\documentclass[%
 reprint,
 amsmath,amssymb,
 aps,
]{revtex4-2}

\usepackage{graphicx}
\usepackage{dcolumn}
\usepackage{bm}

\begin{document}

\preprint{APS/123-QED}
\title{Proposal for constraining non-Newtonian gravity at nm range via criticality enhanced measurement of resonance frequency shift}


\author{Lei Chen}
 \altaffiliation[Also at ]{Key Laboratory of Artificial Structures and Quantum Control (Ministry of Education), School of Physics and Astronomy, Shanghai Jiao Tong University, 800 Dong Chuan Road, Shanghai 200240, China}
 \email{chenleiquantum99@sjtu.edu.cn}
\author{Jian Liu}
 \altaffiliation[Also at ]{Key Laboratory of Artificial Structures and Quantum Control (Ministry of Education), School of Physics and Astronomy, Shanghai Jiao Tong University, 800 Dong Chuan Road, Shanghai 200240, China}
\author{Ka-di Zhu}%
\affiliation{%
 Key Laboratory of Artificial Structures and Quantum Control (Ministry of Education), School of Physics and Astronomy, Shanghai Jiao Tong University, 800 Dong Chuan Road, Shanghai 200240, China
}%





\begin{abstract}
We propose a quantum mechanical method of constraining non-Newtonian gravity at the nanometer range. In this method, a hybrid electro-optomechanical  system is employed. Applying a strong driving field, we can obtain normal mode splitting of the  electromechanical subsystem which is related to the resonance frequency of the mechanical oscillator.
Moreover, we investigate the relationship between the variance of normal mode splitting and the resonance frequency shift induced by the gradient of exotic forces provided that our system is operated at critical points. Furthermore, via suppressing
the Casimir background, we set a constraint on the non-Newtonian gravity which improves the previous bounds by about
a factor of 7 at 1 nanometer range. Our results indicate that our method could be put into consideration in relevant experimental
searches.
\end{abstract}

\maketitle


\section{\label{sec:level1}Introduction}
Though gravity is well described by the Newtonian inverse-square law  in the nonrelativistic limit in a weak gravitational field, it is poorly characterized in the short range \cite{PhysRevLett.124.051301}. In this range, the gravitational potential  between two masses $m_1$ and $m_2$ separated by distance $r$ can be modified as Yukawa potential
\begin{align}
	V_{Yu}(r)=-G\frac{m_1 m_2}r(1+\alpha e^{-r/\lambda}),
\end{align}
where $G$ is the Newtonian gravitational constant, $\alpha$ is the strength of any new interaction, $\lambda=\hbar/m_b c$ is the interaction range, and $m_b$ is the mass of the exchanged boson.
Due to the requirement of unifying gravity and particle physics, solving the cosmological-constant problem etc.\cite{doi:10.1146/annurev.nucl.53.041002.110503}, amounts of short-range gravity experiments
\cite{J2003New,Savas2003Probing,2003Testing,article,2005Constraints,2008Improved,2008Stronger,Masuda2009Limits,2010Advance,2013Constraints,Kamiya2015Constraints,Y2016Stronger,2017Probing,2020Constraints,PhysRevLett.124.101101,PhysRevLett.124.051301}
have been conducted. So far, for the purposes of constraining non-Newtonian gravity and testing  the gravitational inverse square law at short ranges,
many experimental methods have been developed and various kinds of device have been put into use \cite{doi:10.1146/annurev.nucl.53.041002.110503,2009Tests,Murata2015A}.
However, the Yukawa interaction also called non-Newtonian gravity can be well constrained only down to the submillimeter range \cite{PhysRevLett.124.051301}, and it at the ultrashort ranges still needs to be investigated.

In this paper, we develop a quantum mechanical method to constrain Yukawa interaction at short ranges. In our method, a hybrid system consisting of a mechanical oscillator, an optical cavity and a microwave resonator is put into use.  Via driving the microwave resonator with a strong field, we can attain the splitting of normal modes of the electromechanical subsystem. The gradient of  exotic forces would induce the resonance frequency  shift of the mechanical oscillator, resulting in the variance of the normal mode splitting. Based on the relationship between the frequency shift and the variance of splitting,
, we establish our detection principles. Furthermore, we demonstrate how $G$ criticality enhances our detection. Via calculation and reasonable estimation, we set a  constraint on the non-Newtonian gravity which is most stringent at about $3\times {10}^{-10}m<\lambda<5\times{10}^{-8}m$. Finally, we hope our method would be realized experimentally in the near future.

The remainder of the paper is organized as follows: In Sec. II we present our theoretical model, in Sec. III we propose  the detection principles,  in Sec. IV we
summarize the paper. In addition, there are two subsections in Sec. III: in the first subsection we focus on the measurement of resonance frequency shift, in the second one we set a constraint.

\section{\label{sec:level1}Theoretical model}

\begin{figure}
	\includegraphics[width=23em]{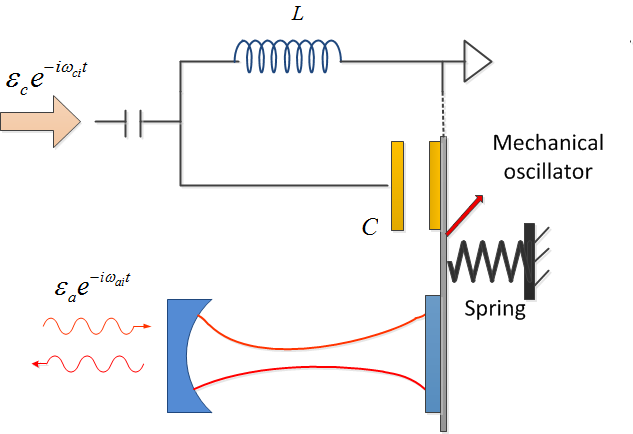}
	\caption{\label{fig:epsart} Schematic diagram of the proposed system. A mechanical oscillator is simultaneously coupled to an optical cavity and a microwave LC resonator . }
\end{figure}

\begin{figure*}
	\includegraphics[width=43em]{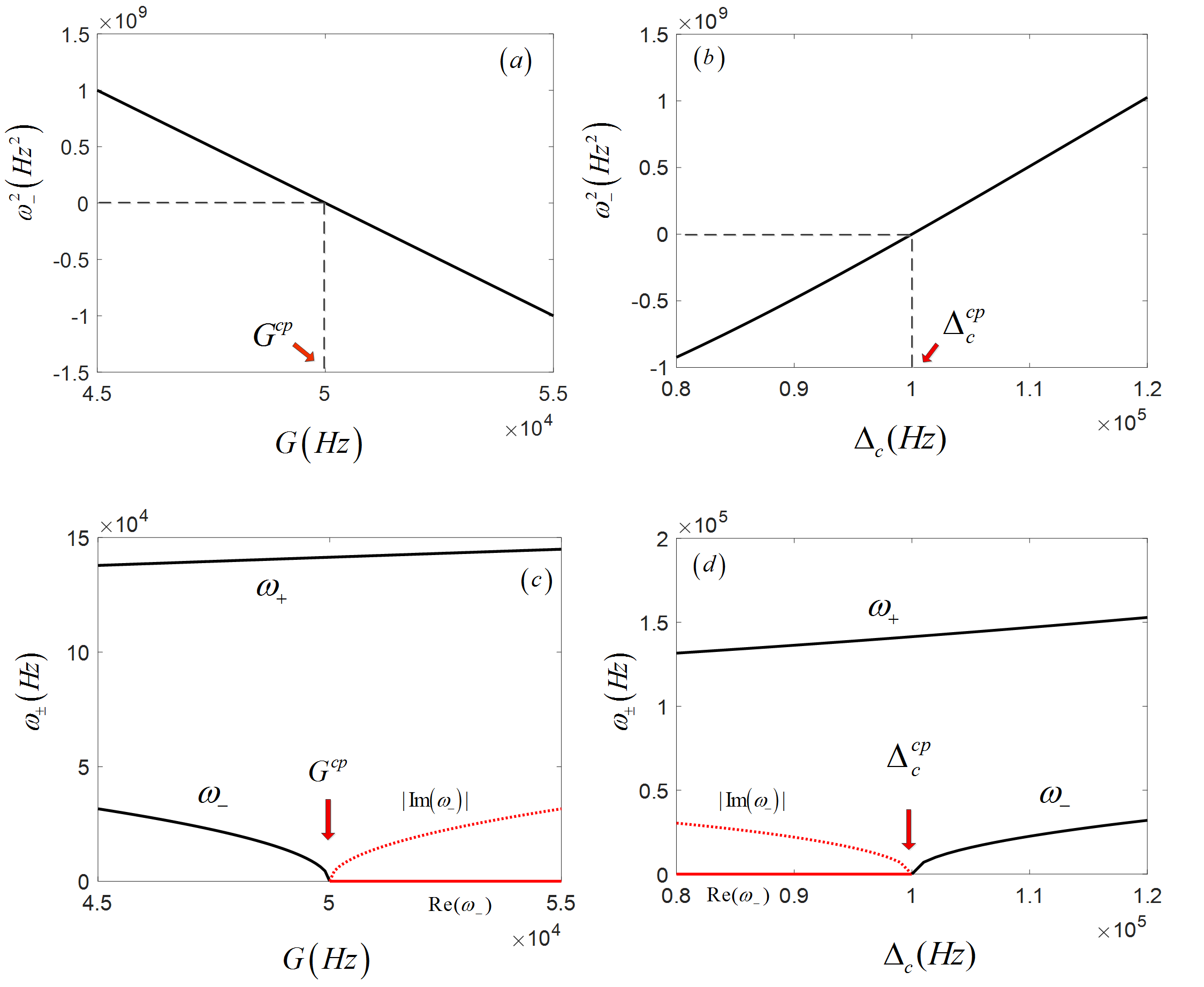}
	\caption{\label{fig:epsart} Criticality of the electromechanical subsystem. (a,c) $\omega_-^2$ and $\omega_{\pm}$ as functions of $G$. As $G$ crosses the critical point $G^{cp}$, $\omega_-$ changes from real to pure imaginary. The parameters used are $\omega_b =10^5 Hz$ and $\Delta_c /\omega_b =1$. (b,d) $\omega_-^2$ and $\omega_{\pm}$ as functions of $\Delta_c$. $\omega_-$ changes from pure imaginary to real when $\Delta_c$ crosses the critical point $\Delta_c^{cp}$. Here $\omega_b $ still takes the value of $10^5Hz$ and the linearized electromechanical coupling strength is $G/\omega_b =0.5$.  }
\end{figure*}
We consider a hybrid electro-optomechanical system. In this system, a mechanical oscillator is coupled to both an optical cavity and a microwave resonator. The microwave resonator is driven by a strong field with amplitude  $\varepsilon_c$ and frequency $\omega_{ci}$ , where $\varepsilon_c$ is related to  the input microwave power $P$ and microwave decay rate $\kappa_c$ by $|\varepsilon_c|=\sqrt{2P\kappa_c/\hbar \omega_{ci}}$. In a frame rotating with frequency $\omega_{ci}$, the Hamiltonian of our system can be described as
\begin{align}
	\hat H /\hbar=&\delta_c {\hat c}^+ \hat c +\omega_a {\hat a}^+ \hat a +\omega_b {\hat b}^+ \hat b
	+g_a {\hat a}^+ \hat a ({\hat b}^+ +\hat b)\notag\\&+g_c {\hat c}^+ \hat c ({\hat b}^+ + \hat b) + i\varepsilon_c ({\hat c}^+ - \hat c),
\end{align}
where the detuning $\delta_c =\omega_c - \omega_{ci}$ and the microwave frequency $\omega_c =1/\sqrt{LC}$, $g_a (g_c)$ denotes the optomechanical (electromechanical) coupling strength at the single-photon level, and $\hat a$ ($\hat b$ or $\hat c$) is the annihilation operator of the optical cavity (the mechanical oscillator or the microwave resonator).
Since the coherent driving is strong, the dynamics of our system can generally be well approximated by a linearised description\cite{Bowen2015Quantum}. According to \cite{PhysRevA.47.3173,PhysRevLett.98.030405,Wilson2007Theory}, Eq. (2) can be transformed into
\begin{align}
	{\hat H}_{lin} /\hbar=&\Delta_c {\hat c}^+ \hat c +{\tilde \omega}_a {\hat a}^+ \hat a +\omega_b {\hat b}^+ \hat b
	+g_a {\hat a}^+ \hat a ({\hat b}^+ +\hat b)\notag\\&-G({\hat c}^+ + \hat c)({\hat b}^+ + \hat b),
\end{align}
with
\begin{subequations}
	\begin{align}
		G=&g_c \sqrt{\frac{2P\kappa_c}{\hbar(\omega_c-\delta_c)(\kappa_c^2+\Delta^2_c)}},\\
		\Delta_c=&\delta_c-\frac{4g^2_c P \kappa_c}{\hbar \omega_b(\omega_c-\delta_c)(\kappa_c^2+\Delta^2_c)},\\
		{\tilde \omega}_a=&\omega_a-\frac{4g_a g_c P \kappa_c}{\hbar \omega_b(\omega_c-\delta_c)(\kappa_c^2+\Delta^2_c)},
	\end{align}
\end{subequations}
where $G$ is the linearized electromechanical coupling strength, $\Delta_c$ is the effective microwave detuning, and ${\tilde \omega}_a$ is the redefined optical frequency.

For the purpose of demonstrating the criticality in the electromechanical subsystem, we employ the method used in \cite{L2013Quantum} and diagonalize this subsystem. As a result, the Hamiltonian ${\hat H}_{lin}$ becomes
\begin{align}
	{\hat H}_{dia} /\hbar=&\omega_- {\hat B}^+ _- {\hat B}_- + \omega_+ {\hat B}^+ _+ {\hat B}_+  +{\tilde \omega}_a {\hat a}^+ \hat a \notag\\
	&+g_- {\hat a}^+ \hat a ({\hat B}^+_- +{\hat B}_-) + g_+ {\hat a}^+ \hat a ({\hat B}^+_+ +{\hat B}_+),
\end{align}
with
\begin{align}
	\omega_{\pm}^2=\frac{1}{2} \left(\Delta_c^2+\omega_b^2 \pm \sqrt{(\omega^2_b -\Delta_c^2)^2 +16 G^2 \Delta_c \omega_b}\right),
\end{align}
where $\omega_{\pm}$ are the normal mode frequencies of the subsystem, $g_{\pm}$ are the effective coupling strengths between the optical photon and the normal modes.

From Eq.(6), we derive $\omega_-=0$ if
\begin{equation}
	\Delta_c \omega_b =4G^2.
\end{equation}
Based this equation, we define $G^{cp} =\sqrt{\Delta_c \omega_b} /2$ and $\Delta^{cp}_c =\frac{4G^2}{\omega_b}$. Obviously, when the values of $\Delta_c$ and $\omega_b$ are fixed and $G$ increases from $G^{cp} =\sqrt{\Delta_c \omega_b} /2$, $\omega_-^2$ would change from zero to negative, as shown in Fig. 2(a). Similarly, when the values of $G$ and $\omega_b$ are specified and $\Delta_c$ varies from $\Delta^{cp}_c $, $\omega_-^2$ possibly changes from zero to negative, as shown in Fig. 2(b). The decrease of $\omega_-^2$ here corresponds to a critical behavior \cite{Sudhir2012Critical}. Along with the decrease of $\omega_-^2$ , the normal mode $\omega_-$ is a standard harmonic oscillator at first ($\omega_-^2 >0$). Then it does not have a bound spectrum and is a free particle($\omega_-^2 =0$), and finally is dynamically unstable ($\omega_-^2 <0$). Note that $\omega_-^2 <0$ means that $\omega_-$ is imaginary. Figure 2. (c) and (d) illustrate the variance of $\omega_{\pm}$. Normal mode splitting is defined as
\begin{equation}
	d=\omega_+ -\omega_-.
\end{equation}
Combining Eq. (6) and (8), we derive
\begin{align}
	d=&\sqrt{\frac{1}{2} \left(\Delta_c^2+\omega_b^2 +\sqrt{(\omega_b^2 -\Delta_c^2 )^2 +16 G^2 \Delta_c \omega_b}\right)}\notag\\&-\sqrt{\frac{1}{2} \left(\Delta_c^2+\omega_b^2 -\sqrt{(\omega_b^2 -\Delta_c^2 )^2 +16 G^2 \Delta_c \omega_b}\right)}.
\end{align}
Till now, our theoretical model has been established. Then we propose our  principles of detecting non-Newtonian gravity in the following.

\section{\label{sec:level1}The detection principles}
\begin{figure}
	\includegraphics[width=23em]{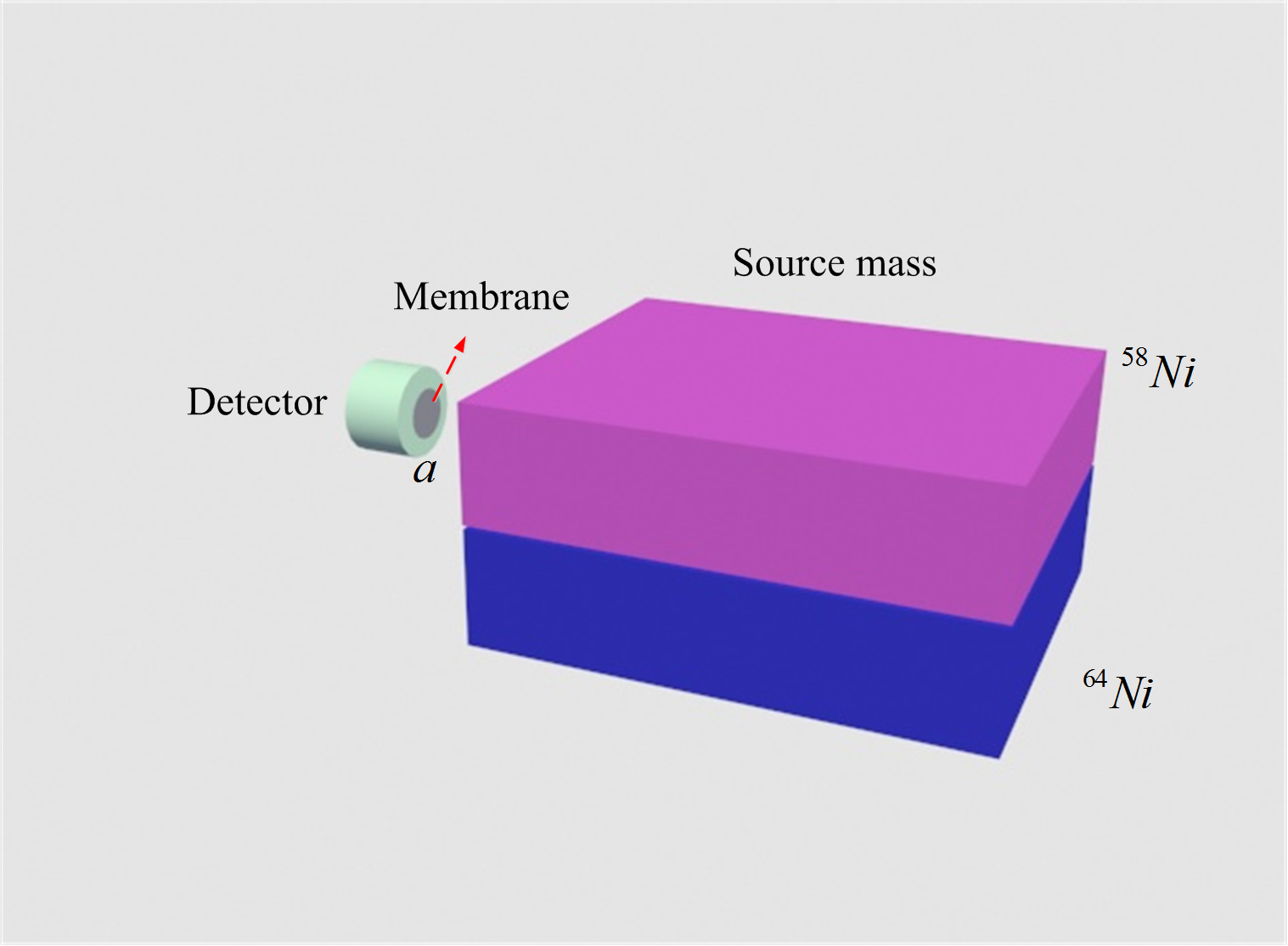}
	\caption{\label{fig:epsart}The setup for the detection of non-Newtonian gravity. A detector with a membrane located in one end  is placed near to the source mass with a separation $a$. The source mass is composed of two blocks which are made of ${}^{58}Ni$ and ${}^{64}Ni$ respectively.  }
\end{figure}
Here we design a micro-size detector whose internal structure is based on the system in Fig. 1. In this detector,  a 1nm thick membrane as seen in Fig. 3 plays the role of the mechanical oscillator in that system. Two  centimeter-scale blocks made of two isotopes of nickel, i.e., ${}^{58}Ni$ and ${}^{64}Ni$ respectively constitute the source mass. As shown in Fig. 3 when the membrane is separated from  the source mass with a distance $a\sim 5nm$, some exotic forces such as Casimir force may exist between the two. The total force gradient from the source mass acting on the membrane modifies its resonance frequency. According to \cite{2012Gradient,2019Precision,Franz2003Advances}, we have
\begin{equation}
	\frac{\delta \omega_b}{\omega_b} = -\frac1{2m_b\omega_b^2} \frac{\partial{F_{total}(a)}}{\partial{a}},
\end{equation}
where the frequency shift is $\delta \omega_b =\omega_b^{\prime} -\omega_b $, $\omega_b^{\prime}$ is the modified resonance frequency, $m_b$ is the mass of the membrane, and $F_{total}(a)$ is the total force .
$|F_{total}(a)|$  generally decreases as $a$ increases, resulting $\delta \omega_b <0$ and $\omega_b^{\prime} <\omega_b$. From Eq. (9) we find that the variance of the resonance frequency $\omega_b$ would induce the variation of normal mode splitting $d$ provided that $G$ and $\Delta_c$ are fixed. Based on this, considering the  criticality as shown in Fig. (2),  we establish a method to measure the force gradient induced variance of $\omega_b$ and then set a prospective constraint on non-Newtonian gravity.

\subsection{\label{sec:level2}Measurement of resonance frequency shift enhanced by the $G$ criticality}

\begin{figure}
	\includegraphics[width=23em]{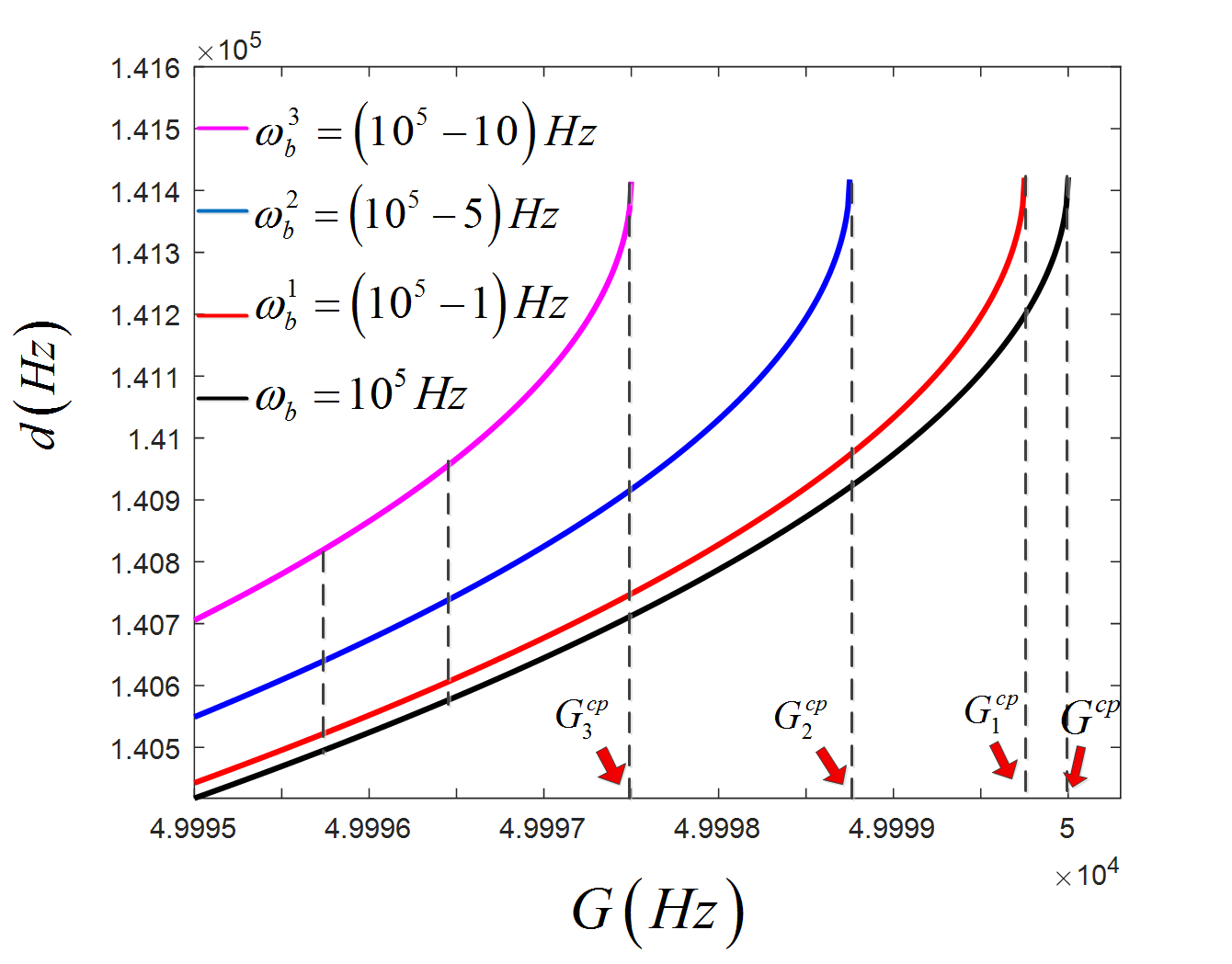}
	\caption{\label{fig:epsart}$d$ as a function of $G$. In the four curves, resonance frequency takes values of $\omega_b$ and $\omega_b^i$ ($i=1,2,3$) and the rightmost values of  $G$ are all the critical points ($ G^{cp}$ and $G^{cp}_i ,i=1,2,3$ ).  }
\end{figure}

\begin{figure*}
	\includegraphics[width=45em]{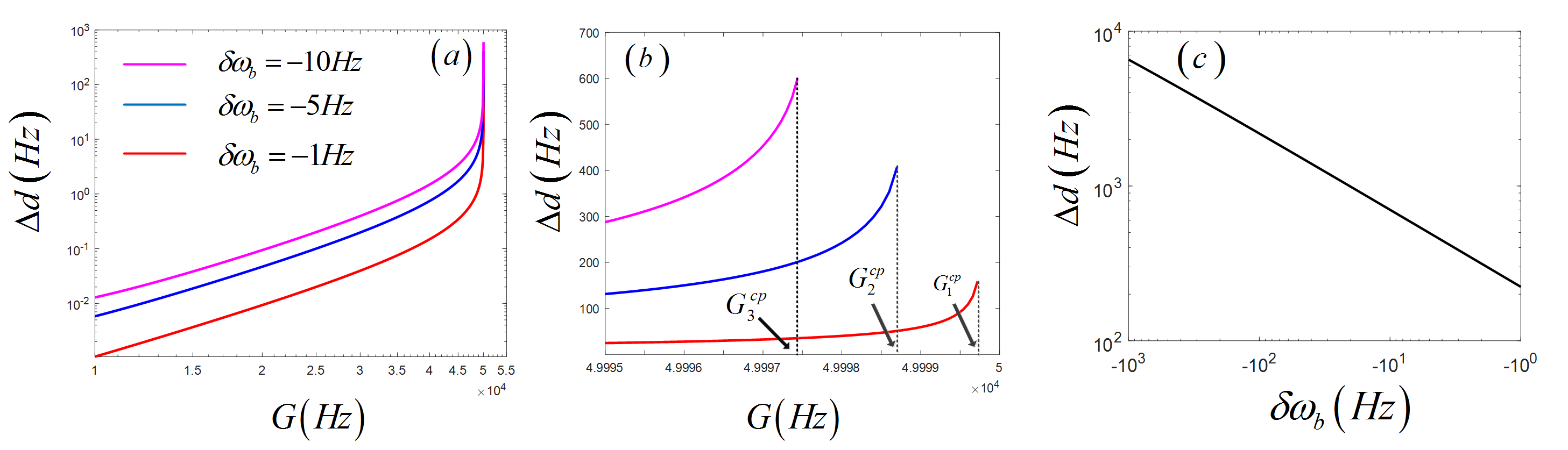}
	\caption{\label{fig:epsart}$G$ criticality enhanced measurement. (a)-(b) $\Delta d$ as a function of $G$. In the three curves, the resonance frequency shift $\delta \omega_b$ takes values of -1,-5, -10Hz respectively and the rightmost values of  $G$ are all the corresponding critical points ($G^{cp}_i (i=1,2,3)$ ). (c)$\Delta d$ is the function of $\delta \omega_b$ according to Eqs. (13)-(14).   }
\end{figure*}

Here the parameters used are $\omega_b =10^5 Hz$ and $\Delta_c ={10}^5Hz$, just the same as Fig. 2(a,c). Different from these
two parameters, the value of $G$ is variable.
The critical point is $G^{cp}=50000Hz$.
The possible modified frequencies $\omega_b^i$ (for $i=1,2,3$) are set as $\omega_b^1 =({10}^{5} -1)Hz, \omega_b^2 =({10}^{5} -5)Hz,$ and $\omega_b^3 =({10}^{5} -10)Hz$ respectively.  From Eq. (9), it is seen that if the resonance frequency takes values of $\omega_b$ or $\omega_b^i$ ($i=1,2,3$), $d$ would be functions of $G$ , which are plotted in Fig. 4.
In these four curves we choose four critical points as the maximum values of $G$ to make sure $d$ is real.
Since it is defined that $G^{cp} =\sqrt{\Delta_c \omega_b} /2$, we can obtain the corresponding $G^{cp}_i$ (for $i=1,2,3$) as $G_1^{cp} =49999.75Hz, G_2^{cp} =49998.75Hz,$ and $G_3^{cp} =49997.50Hz.$

From Fig. 4 we see that if resonance frequency is modified from  to $\omega_b =10^5Hz$ to $\omega_b^i$ (for $i=1,2,3$), normal mode splitting $d$ would shift. Furthermore, it seems that the three shifts corresponding to three modifications of the resonance frequency all reach their maximum at the corresponding critical points. we assume that $d\to d+\Delta d$ corresponds to $\omega_b\to\omega_b+\delta\omega_b$. The relationship between $\Delta d$ and $\delta\omega_b$ can be expressed as

\begin{align}
	\Delta d= &\sqrt{\frac{1}{2} (u_\delta + \sqrt {v_\delta})} - \sqrt{\frac{1}{2} (u_\delta -\sqrt {v_\delta})}\notag\\&- \sqrt{\frac{1}{2} (u + \sqrt {v})} +\sqrt{\frac{1}{2} (u - \sqrt {v})},
\end{align}

where \begin{align}
	u&=\Delta_c^2 + \omega_b^2, \notag\\ v&=(\omega_b^2 -\Delta_c^2 )^2+16 G^2 \Delta_c \omega_b,\notag\\u_\delta &=\Delta_c^2 + (\omega_b+\delta\omega_b)^2, \notag\\ v_\delta&=[(\omega_b+\delta\omega_b)^2 -\Delta_c^2 ]^2+16 G^2 \Delta_c (\omega_b+\delta\omega_b).
\end{align}
Since the values of $\omega_b$ and $\Delta_c$ have been specified, if the values of resonance frequency shift $\delta\omega_b$ are set as -1, -5, -10Hz respectively, the  value of normal mode splitting shift $\Delta d$ would be functions of $G$
according to Eqs. (11) and (12). Note that these three values of $\delta\omega_b$ correspond to $\omega_b^i$ ($i=1,2,3$) respectively. These three functions where $G$ takes values as $ {10}^4 Hz \le G \le G^{cp}_i (i=1,2,3)$  are plotted
in Fig. 5(a) and (b). In Fig. 5(b) the functions are plotted only when the values of $G$ which are very close to three corresponding critical points. And it is complementary to Fig. 5(a). From Fig. 5(a) and (b) we can find that for all three modifications of resonant frequency the value of $\Delta d$ at critical point is several orders bigger than it at $G={10}^4 Hz$.
Generally speaking, utilizing $G$ criticality can enhance the measurement of resonance frequency shift. Then we investigate how  $\Delta d$ is dependent on  $\delta\omega_b$ if $G$ takes value of the relating critical points.

We define $G^{cp}_{\delta} =\sqrt{\Delta_c (\omega_b + \delta\omega_b)} /2$. Consequently $G^{cp}_{\delta}$ is the critical point corresponding to $\omega_b + \delta\omega_b$.  In Eq. (9) We substitute $G^{cp}_{\delta}$ for $G$, and derive

\begin{align}
	\Delta d= \sqrt{\frac{1}{2} (u_\delta + \sqrt {v_\delta^G})} - \sqrt{\frac{1}{2} (u + \sqrt {v^G})} +\sqrt{\frac{1}{2} (u - \sqrt {v^G})},
\end{align}

where \begin{align}
	v^G&=(\omega_b^2 -\Delta_c^2 )^2+16 (G^{cp}_{\delta})^2 \Delta_c \omega_b,\notag\\ v_\delta^G&=[(\omega_b+\delta\omega_b)^2 -\Delta_c^2 ]^2+16 (G^{cp}_{\delta})^2 \Delta_c (\omega_b+\delta\omega_b),
\end{align}
and $u_{\delta}$ and $u$ have been defined in the above.   The relationship between $\Delta d$ and $\delta\omega_b$ is shown in Fig. 5(c). Till now, we have demonstrated how the $G$ criticality enhances the measurement of resonance frequency shift. Furthermore, this measurement can be enhanced by the $\Delta_c$ criticality in a similar way.
\subsection{\label{sec:level2}A  constraint for the non-Newtonian gravity}

\begin{figure}[b]
	\includegraphics[width=25em]{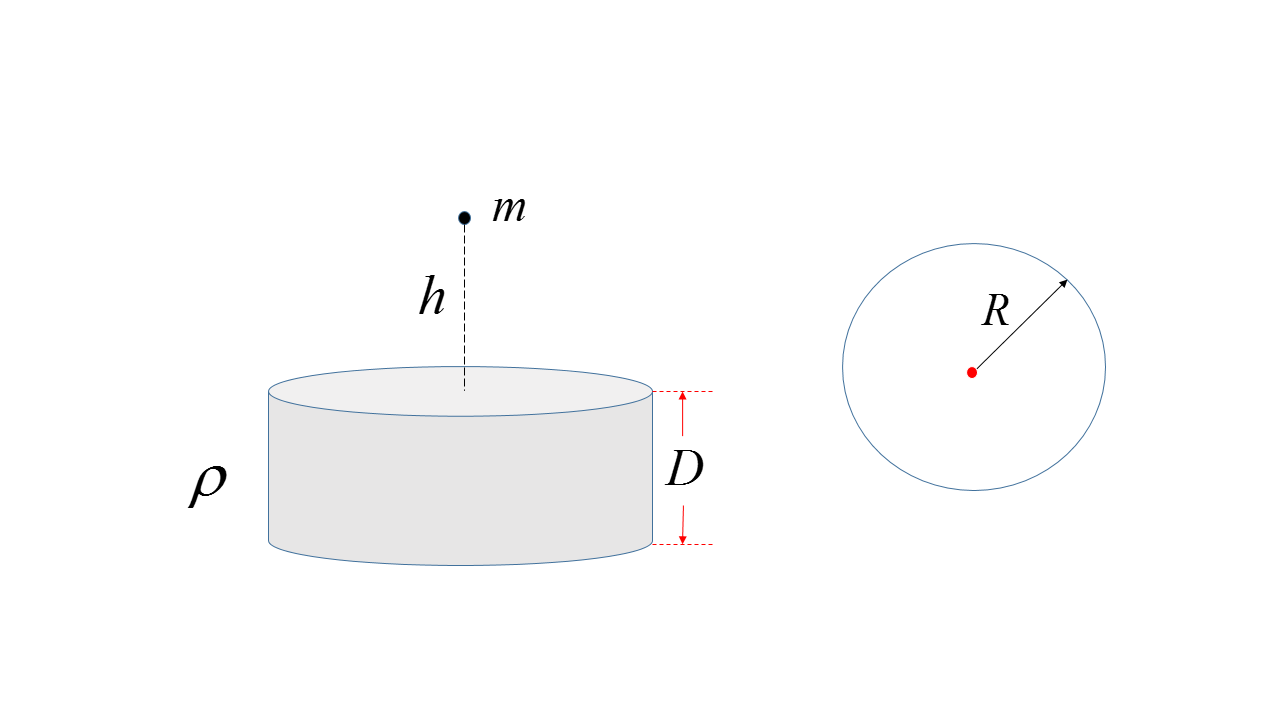}
	\caption{\label{fig:epsart}A particle with mass $m$ is located at a distance $h$ above a cylinder with density $\rho$, radius $R$, and thickness $D$.   }
\end{figure}

In this section, we develop a method of constraining  the non-Newtonian gravity between the membrane and the source mass.
Since Casimir forces depend to a good approximation on the electronic properties of materials while gravitational interaction involve couplings to both electrons and nucleons \cite{2003Testing}, using materials with very similar properties can be considered in order to suppress the Casimir background. And two isotopes of nickel, i.e., ${}^{58}Ni$ and ${}^{64}Ni$ , are adopted in our setup to eliminate the Casimir effect. The detector can move along the surface of the source mass provided that the separation between the two is constant. In the following ,we derive  the non-Newtonian gravity at first.

We consider a simple case where a particle $m$ is located at the axis of symmetry of a cylinder with a separation $h$ as shown in Fig. 5. The radius , density, and thickness of this cylinder are signified by $R$, $\rho$, $D$ respectively. Provided $R\gg D$ and $R\gg h$, by integrating Eq. (1) over the volume of the cylinder, we obtain the Yukawa energy
\begin{align}
	E_{Yu}(h)\approx &-Gm\rho2\pi RD \notag\\& + Gm\rho\alpha2\pi\lambda[e^{-\frac R \lambda}D-\lambda e^{-\frac h \lambda}(1-e^{-\frac D \lambda})] .
\end{align}
Calculating the negative derivative of (15) with respect to $h$,  we finds  non-Newtonian gravity acting on the particle
\begin{align}
	F_{G}(h)=-\frac {d E_{Yu}(h)}{dh }=  -e^{-\frac h \lambda} Gm\rho\alpha2\pi\lambda(1-e^{-\frac D \lambda})  .
\end{align}
Considering the sizes of the detector and two blocks, we conclude that with the assumption $\lambda \le 1\mu m$, the  non-Newtonian gravity can be expressed as
\begin{align}
	F^{r}_{G}(a)\approx &-e^{-\frac a \lambda} Gm_b\rho_{r}\alpha2\pi\lambda,\notag\\
	F^{b}_{G}(a)\approx&-e^{-\frac a \lambda} Gm_b\rho_{b}\alpha2\pi\lambda,
\end{align}
where $F^{r}_{G}(a)$ corresponds to  case I in which detector near the red block, $F^{b}_{G}(a)$ corresponds to  case II in which detector near the red one, and $\rho_{r}$ and $\rho_{b}$ denote the density of red block (${}^{58}Ni$) and the one of blue block (${}^{64}Ni$) respectively.
Considering these two cases, Eq. (10) can be rewritten as
\begin{align}
	\frac{\delta^r _{\omega_b}}{\omega_b} = -\frac1{2m_b\omega_b^2} \frac{\partial{F^r_{total}(a)}}{\partial{a}},\notag\\
	\frac{\delta^b _{\omega_b}}{\omega_b} = -\frac1{2m_b\omega_b^2} \frac{\partial{F^b_{total}(a)}}{\partial{a}},
\end{align}
where $\delta^r _{\omega_b}, F^r_{total}(a)$ and $\delta^b _{\omega_b}, F^b_{total}(a)$ correspond to case I and II respectively.
Since two isotopes of nickel are put into use, we can obtain
\begin{equation}
	F^r_{total}(a)-F^b_{total}(a)\approx F^{r}_{G}(a)-F^{b}_{G}(a).
\end{equation}
From Eqs. (17)-(19), we derive that
\begin{equation}
	\alpha= \frac {\omega_b(\delta^r _{\omega_b}-\delta^b _{\omega_b}) } {G\pi(\rho_{b}-\rho_{r})}e^{\frac a \lambda}.
\end{equation}
Since $\rho_{b}-\rho_{r}>0$, we can obtain that
\begin{equation}
	|\alpha|= \frac {\omega_b|\delta^r _{\omega_b}-\delta^b _{\omega_b}| } {G\pi(\rho_{b}-\rho_{r})}e^{\frac a \lambda}.
\end{equation}

For the purpose of setting a constraint on $|\alpha|$, we need to determine the minimum detectable  value of $|\delta^r _{\omega_b}-\delta^b _{\omega_b}|$ signified by $|\delta^r _{\omega_b}-\delta^b _{\omega_b}|_m$. Further, we can assume that
\begin{equation}
	|\delta^r _{\omega_b}-\delta^b _{\omega_b}|_m \approx |\delta {\omega_b}|_m,
\end{equation}
where $|\delta {\omega_b}|_m $denotes the minimum detectable  value of $|\delta {\omega_b}|$. Provided that $G$ takes value of the relating critical points, $|\delta {\omega_b}|_m $ can be associated with the minimum distinguishable value of $\Delta d$ signified by $\Delta d_m$ via Eqs. (13)-(14), where $\Delta d$  and $\delta {\omega_b}$ are substituted by $\Delta d_m$ and
$-|\delta {\omega_b}|_m $ respectively. Now we demonstrate how to determine $\Delta d_m$. We focus on the article titled "Observation of strong coupling between a micromechanical resonator and an optical cavity field"\cite{2009Natureletters}. In it, the observation of optomechanical normal mode splitting is reported. In Fig. 2(b) of this article, we find that there is minor difference between theory and experimental data. Based on this difference, we estimate that
\begin{equation}
	\Delta d_m \approx 0.01 \omega_b.
\end{equation}
Since $\omega_b={10}^5 Hz$, we obtain $\Delta d_m \approx {10}^3 Hz$.  Further, according to Fig. 5(c), we estimate that $|\delta {\omega_b}|_m \sim 10Hz$. Then according to Eq. (22)it is attained that $|\delta^r _{\omega_b}-\delta^b _{\omega_b}|_m\sim 10 Hz$. Our constraints on $|\alpha|$ are set as shown in Fig. 7 using
\begin{equation}
	|\alpha|= \frac {\omega_b|\delta^r _{\omega_b}-\delta^b _{\omega_b}|_m } {G\pi(\rho_{b}-\rho_{r})}e^{\frac a \lambda},
\end{equation}
where $|\delta^r _{\omega_b}-\delta^b _{\omega_b}|_m$ takes value of $10Hz$, and the value of $\lambda$ meets the assumption $\lambda \le 1\mu m$.
\begin{figure}[b]
	\includegraphics[width=25em]{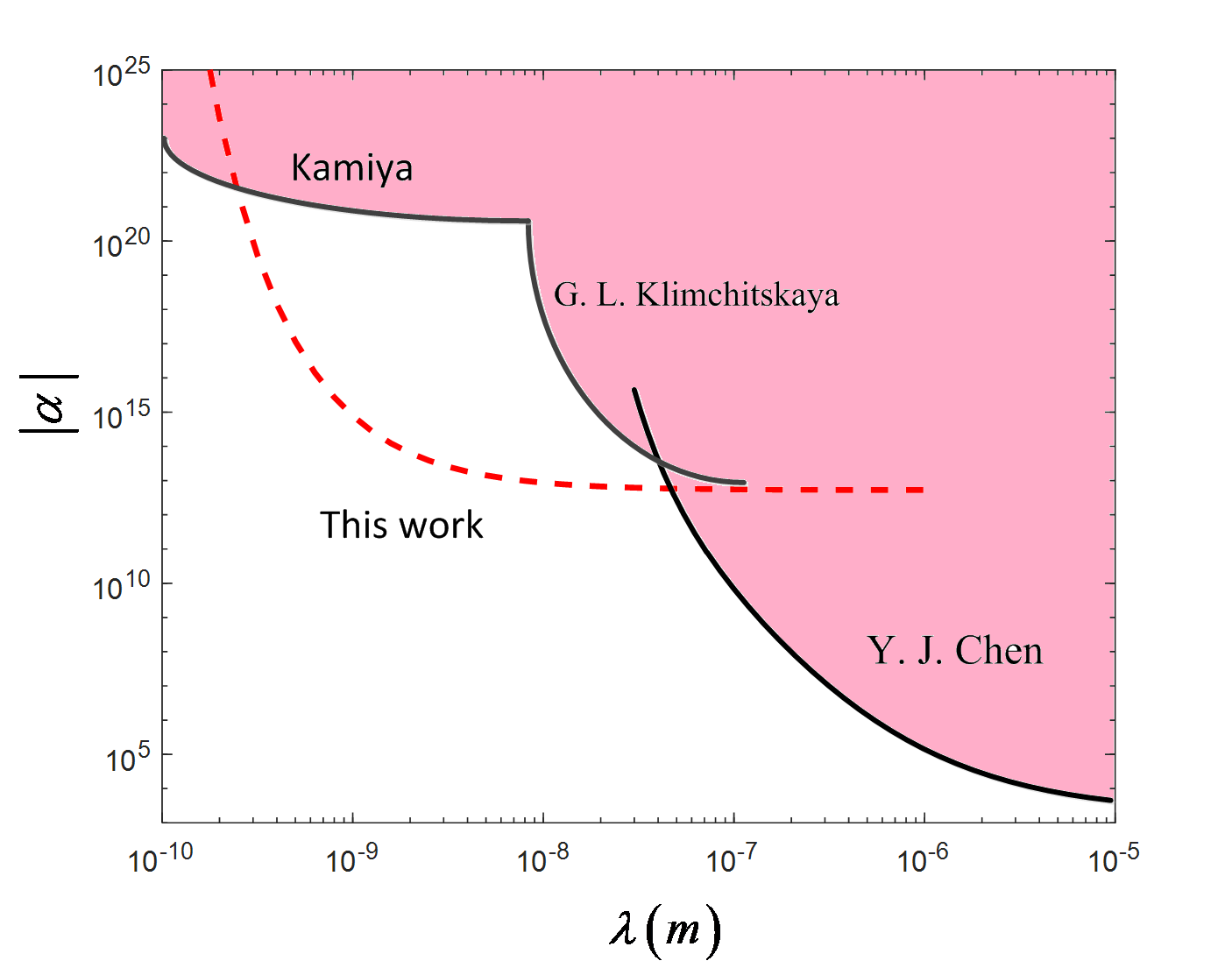}
	\caption{\label{fig:epsart}The $|\alpha|-\lambda$ plot for constraints established by Y.J.Chen et al., Klimchitskaya et al., Kamiya et al., and our work respectively. The pink region is excluded.  }
\end{figure}

Now we focus on Fig. 7. Y.J.Chen establish a upper bound in the $40-8000$ $ nm $ length scale bsed on differential force measurements between a test mass and rotating source masses \cite{Y2016Stronger}. Klimchitskaya et al. set a upper bound approximately at ${10}^{-8}m<\lambda<2\times{10}^{-7}m$ \cite{2013Constraints}. Kamiya et al. provide constraints in the Nanometer Range by performing a neutron scattering experiment \cite{Kamiya2015Constraints}. Our constraints represented by the red dashed curve are most stringent at about $3\times {10}^{-10}m<\lambda<5\times{10}^{-8}m$. The pink region is excluded.
\\

\section{\label{sec:level1}Discussion and conclusion}
In sum, we have proposed a quantum mechanical method of constraining non-Newtonian gravity with a hybrid electro-optomechanical system. By employing the source mass consisting of  two isotopes of nickel in order to suppress the Casimir background, via $G$  criticality enhanced measurement of resonance frequency shift, we can detect and constrain the non-Newtonian gravity. Based on the experimental results relating to normal mode splitting, we set a constraint which improves the previous bounds by about a factor of 7 at $\lambda=1nm$.  We hope our work can enrich the experimental methods of searching for non-Newtonian gravity and promote the searches for this exotic interaction at the nanometer range.

\begin{acknowledgments}
This work was supported by Natural Science Foundation of Shanghai (No. 20ZR1429900).
\end{acknowledgments}

\bibliography{apssamp}

\end{document}